\documentclass[12pt,onecolumn]{revtex4}
\usepackage{amssymb}

\begin{document}

\title{Asymptotically (anti)-de Sitter solutions in Gauss-Bonnet gravity without a
cosmological constant}
\author{M. H. Dehghani}\email{mhd@shirazu.ac.ir}
\address{Physics Department and Biruni Observatory, Shiraz University, Shiraz 71454, Iran;\\  Institute for Studies in Theoretical Physics and Mathematics (IPM) P.O. Box 19395-5531, Tehran, Iran\\and\\
         Research Institute for Astrophysics and Astronomy of Maragha (RIAAM), Maragha, Iran}

\begin{abstract}
In this paper we show that one can have asymptotically de Sitter
(dS), anti-de Sitter (AdS) and flat solutions in Gauss-Bonnet
gravity without any need to a cosmological constant term in field
equations. First, we introduce static solutions whose $3$-surfaces
at fixed $r$ and $t$ have constant positive ($k=1$), negative
($k=-1$), or zero ($k=0$) curvature. We show that for $k=\pm1$,
one can have asymptotically dS, AdS and flat spacetimes, while for
the case of $k=0$, one has only asymptotically AdS solutions. Some
of these solutions present naked singularities, while some others
are black hole or topological black hole solutions. We also find
that the geometrical mass of these $5$-dimensional spacetimes is
$m+2\alpha \left| k\right| $, which is different from the
geometrical mass, $m $, of the
solutions of Einstein gravity. This feature occurs only for the $5$%
-dimensional solutions, and is not repeated for the solutions of
Gauss-Bonnet gravity in higher dimensions. We also add angular
momentum to the static solutions with $k=0$, and introduce the
asymptotically AdS charged rotating solutions of Gauss-Bonnet
gravity. Finally, we introduce a class of solutions which yields
an asymptotically AdS spacetime with a longitudinal magnetic field
which presents a naked singularity, and generalize it to the case
of magnetic rotating solutions with two rotation parameters.
\end{abstract}

\maketitle


\section{Introduction}

It seems established that at the present epoch the Universe expands with
acceleration. This follows directly from the observation of high red-shift
supernova \cite{Per} and indirectly from the measurement of angular
fluctuations of cosmic microwave background fluctuations \cite{Lee}. These
astrophysical data have created a great deal of attention to the\
asymptotically de Sitter (dS) spacetimes. On the other hand asymptotically
anti-de Sitter (AdS) spacetimes continue to attract more attention due to
the fact that there is a correspondence between supergravity\ (the
low-energy limit of string theory) in $(n+1)$-dimensional asymptotically AdS
spacetimes and conformal field theory\ (CFT) living on an $n$-dimensional
boundary known as the AdS/CFT correspondence

The simplest way of having an asymptotically (A)dS spacetime is to add a
cosmological constant term to the right hand side of Einstein equation.
However, the cosmological constant meets its well known cosmological, fine
tuning and coincidence problems \cite{Cal}. Thus, it seems natural to get
rid of the cosmological constant and look for an alternative candidate for
it. In the context of classical theory of gravity, the second way of having
an asymptotically (A)dS spacetime is to add higher curvature terms to the
left hand side of Einstein equation. The way that I deal with the
asymptotically (A)dS spacetime is the latter one. Indeed, it seems natural
to reconsider the left hand side of Einstein equation, if one intends to
investigate classical gravity in higher dimensions.

The possibility that spacetime may have more than four dimensions is now a
standard assumption in high energy physics. From a cosmological point of
view, our observable Universe may be viewed as a brane embedded into a
higher dimensional spacetime. The idea of brane cosmology is also consistent
with string theory, which suggests that matter and gauge interaction
(described by an open string) may be localized on a brane, embedded into a
higher dimensional spacetime. The field represented by closed strings, in
particular, gravity, propagate in the whole of spacetime.

This underscores the need to consider gravity in higher dimensions. In this
context one may use another consistent theory of gravity in any dimension
with a more general action. This action may be written, for example, through
the use of string theory. The effect of string theory on classical
gravitational physics is usually investigated by means of a low energy
effective action which describes gravity at the classical level \cite{Wit}.
This effective action consists of the Einstein-Hilbert action plus
curvature-squared terms and higher powers as well, and in general give rise
to fourth order field equations and bring in ghosts. However, if the
effective action contains the higher powers of curvature in particular
combinations, then only second order field equations are produced and
consequently no ghosts arise \cite{Zum}. The effective action obtained by
this argument is precisely of the form proposed by Lovelock \cite{Lov}. The
appearance of higher derivative gravitational terms can be seen also in the
renormalization of quantum field theory in curved spacetime \cite{BDav}.

In this paper we want to restrict ourself to the first two terms
of Lovelock gravity, which are the Einstein-Hilbert and the
Gauss-Bonnet terms. The latter term appears naturally in the
next-to-leading order term of the heterotic string effective
action and plays a fundamental role in Chern-Simons gravitational
theories \cite{Cham}. From a geometric point of view, the
combination of the Einstein-Gauss-Bonnet terms constitutes, for
five-dimensional spacetimes, the most general Lagrangian producing
second order field equations, as in the four-dimensional gravity
where the Einstein-Hilbert action is the most general Lagrangian
producing second order field equations \cite{Lan}. However,
Gauss-Bonnet term is topological in $4$-dimensions, and hence has
no dynamics. Indeed, if this term had made a nontrivial
contribution in 4-dimensions, then it would have conflicted with
the $1/r$ character of the potential because of the presence of
$(\nabla \phi)^4$ terms in the equation \cite{Dad}.

Thus, the problems with the cosmological constant, the need to go
to higher dimensional spacetime, and the interest in
asymptotically (A)dS spacetimes provide a strong motivation for
considering asymptotically (A)dS solutions of the
Einstein-Gauss-Bonnet gravity without cosmological constant.
Recently I introduced a model for Universe in Gauss-Bonnet gravity
without a \ cosmological constant, which is asymptotically de
Sitter. In that model, one does not need to assume any kind of
exotic dark energy, in order to explain the acceleration of the
expanding Universe \cite{Deh3}. Most of the solutions of
Gauss-Bonnet gravity which have been found till now are the
solutions with nonzero cosmological constant. Static spherically
symmetric black hole solutions of the Gauss-Bonnet gravity were
found in Ref. \cite {Des}. Black hole solutions with nontrivial
topology were also studied in Refs. \cite{Cai,Aros,Ish}. The
thermodynamics of charged static spherically symmetric black hole
solutions was considered in \cite{Odin}. All of these known
solutions are static. Recently I introduced two classes of
asymptotically anti-de Sitter rotating solutions in the
Einstein-Gauss-Bonnet gravity and considered their thermodynamics
\cite {Deh1,Deh2}. Also, the linearized gravity on a single de
Sitter brane in Gauss-Bonnet theory has been investigated
\cite{Sas}.

The outline of our paper is as follows. We give a brief review of the field
equations in Sec. \ref{Fiel}. In Sec. \ref{Stat}, we introduce the static
solutions of Gauss-Bonnet gravity without a cosmological constant term in
the presence of electromagnetic field, and show that these solutions
generate asymptotically (anti)-de Sitter and flat spacetimes. In Sec. \ref
{Rot}, we find two classes of asymptotically AdS rotating solutions . We
finish our paper with some concluding remarks.

\section{Field Equations in Gauss-Bonnet Gravity Without a Cosmological
Constant}

\label{Fiel} The most fundamental assumption in standard general relativity
is the requirement that the field equations be generally covariant and
contain at most second order derivative of the metric. Based on this
principle, the most general classical theory of gravitation in five
dimensions is the Einstein-Gauss-Bonnet gravity. The gravitational action of
this theory in five dimensions for the spacetime $(\mathcal{M}$, $g_{\mu \nu
})$ can be written as
\begin{equation}
I_{G}=\frac{1}{2}\int_{\mathcal{M}}dx^{5}\sqrt{-g}[R+\alpha (R_{\mu \nu
\gamma \delta }R^{\mu \nu \gamma \delta }-4R_{\mu \nu }R^{\mu \nu
}+R^{2})+F_{\mu \nu }F^{\mu \nu }],  \label{Ig}
\end{equation}
where $R$, $R_{\mu \nu \rho \sigma }$, and $R_{\mu \nu }$ are the Ricci
scalar and Riemann and Ricci tensors of the spacetime, $F_{\mu \nu
}=\partial _{\mu }A_{\nu }-\partial _{\nu }A_{\mu }$ is the electromagnetic
tensor field, $A_{\mu }$ is the vector potential, and $\alpha $ is the
Gauss-Bonnet coefficient with dimension $(\mathrm{length})^{2}$. Of course,
one may add a constant term to the above Lagrangian, playing the role of
cosmological constant term. But, as we mentioned before, this creates its
own problems and therefore we don't disturb ourself with it. Indeed, here we
want to obtain asymptotically (A)dS solutions without a cosmological
constant term. Varying the action over the metric tensor $g_{\mu \nu }$ and
electromagnetic field $F_{\mu \nu }$, the equations of gravitational and
electromagnetic fields are obtained as

\begin{eqnarray}
&&\ R_{\mu \nu }-\frac{1}{2}g_{\mu \nu }R-\alpha \{\frac{1}{2}g_{\mu \nu
}(R_{\kappa \lambda \rho \sigma }R^{\kappa \lambda \rho \sigma }-4R_{\rho
\sigma }R^{\rho \sigma }+R^{2})  \nonumber \\
&&-2RR_{\mu \nu }+4R_{\mu \lambda }R_{\text{ \ }\nu }^{\lambda }+4R^{\rho
\sigma }R_{\mu \rho \nu \sigma }-2R_{\mu }^{\ \rho \sigma \lambda }R_{\nu
\rho \sigma \lambda }\}=T_{\mu \nu },  \label{Geq}
\end{eqnarray}
\begin{equation}
\nabla _{\mu }F_{\mu \nu }=0,  \label{EMeq}
\end{equation}
where $T_{\mu \nu }$ is the electromagnetic stress tensor
\begin{equation}
T_{\mu \nu }=2{F^{\lambda }\ _{\mu }}F_{\lambda \nu }-\frac{1}{2}F_{\lambda
\sigma }F^{\lambda \sigma }g_{\mu \nu }.  \label{Str}
\end{equation}
Equation (\ref{Geq}) does not contain the derivative of the curvatures, and
therefore the derivatives of the metric higher than two do not appear. Thus,
the Gauss-Bonnet gravity is a special case of higher derivative gravity.

\section{The static solutions}

\label{Stat}

\label{dS}\label{Lon}Here we want to obtain the $5$-dimensional static
solutions of Eq. (\ref{Geq})-(\ref{Str}), which are asymptotically (anti)-de
sitter or flat. We assume that the metric has the following form:
\begin{equation}
ds^{2}=-f(r^{\prime })dt^{2}+\frac{dr^{\prime 2}}{f(r^{\prime })}+r^{\prime
2}d\Omega ^{2},  \label{Met1}
\end{equation}
where $d\Omega ^{2}$ is the metric of a $3$-dimensional hypersurface with
constant curvature $6k$ given as
\begin{eqnarray}
d\Omega ^{2} &=&d\theta ^{2}+\sin ^{2}\theta (d\phi ^{2}+\sin ^{2}\phi d\psi
^{2});\hspace{0.8cm} k=1,  \label{Sph} \\
&=&d\theta ^{2}+\sinh ^{2}\theta (d\phi ^{2}+\sin ^{2}\phi d\psi ^{2});%
\hspace{0.5cm}k=-1,  \label{Hyp} \\
&=&\alpha ^{-1}dx^{2}+\sum_{i=1}^{2}d\phi _{i}^{2}; \hspace{2.8cm}
k=0. \label{flat}
\end{eqnarray}
Note that the coordinates $x$ has the dimension of length, while the angular
coordinates $\theta $, $\phi $, $\psi $ and $\phi _{i}$'s are dimensionless
as usual. The coordinates $\theta $ and $\phi $ lies in the interval $[0,\pi
]$ and $\psi $ and $\phi _{i}$'s range $0\leq \phi _{i}<2\pi $. The
assumption that there exist a charged $q$ at $r=0$ ($q$ is a point charge
for $k=\pm 1$, and is the charge density of a line charge for $k=0$ cases
respectively) means that the vector potential may be written as
\begin{equation}
A_{\mu }=h(r^{\prime })\delta _{\mu }^{0}.  \label{Pot1}
\end{equation}
The functions $f(r^{\prime })$ and $h(r^{\prime })$ may be
obtained by solving the field equations (\ref{Geq})-(\ref{Str}).
Using Eq. (\ref{EMeq}) one obtains
\begin{equation}
r^{\prime }\frac{\partial ^{2}h}{\partial r^{\prime 2}}+3\frac{\partial h}{%
\partial r^{\prime }}=0.  \label{Em1}
\end{equation}
Thus, $h(r^{\prime })=-C_{1}/r^{\prime 2}$, where $C_{1}$ is an
arbitrary real constant. If one uses the Gauss law for the
electric field, then he obtains $2C_{1}=q$. To find the function
$f(r)$, one may use any components of Eq. (\ref{Geq}). The
simplest equation is the $r^{\prime }r^{\prime }$ component of
these equations which can be written as
\begin{equation}
\lbrack 12\alpha r^{\prime 3}(1-f)+3r^{\prime
5}]\frac{df}{dr^{\prime}}-6r^{\prime 4}(1-f)+2q^{2}=0.
\label{rrcom}
\end{equation}
The solutions of Eq. (\ref{rrcom}) can be written as
\begin{equation}
f(r^{\prime })=k+\frac{r^{\prime 2}}{4\alpha }\pm \sqrt{\frac{r^{\prime 4}}{%
16\alpha ^{2}}+\left( \left| k\right| +\frac{m}{2\alpha }\right) -\frac{q^{2}%
}{6\alpha r^{\prime 2}}},  \label{Fg1}
\end{equation}
where $m$ is an arbitrary constant. Also, it is remarkable to note that for
large values of $r$, the function $f(r)$ can be written as:
\begin{equation}
f_{\infty }(r^{\prime })=k+\frac{r^{\prime 2}}{4\alpha }(1\pm 1)\pm \frac{%
m+2\left| k\right| \alpha }{r^{\prime 2}}\mp \frac{q^{2}}{3r^{\prime 4}},
\label{Fginf}
\end{equation}
which shows that the geometrical mass of the spacetime is $m+2\alpha \left|
k\right| $. Thus, the mass of a five-dimensional spacetime in Gauss Bonnet
gravity for $k=\pm 1$, differs from that of Einstein gravity by a term which
is proportional to $6\alpha \left| k\right| $. Note that the Gauss-Bonnet
term decreases the mass of the spacetime for negative $\alpha $, and
increases the mass for positive $\alpha $. It is worthwhile to mention that
this occurs only for the five-dimensional spacetime. For higher-dimensional
solutions in Gauss-Bonnet gravity $(n+1>5)$ the function $f(r)$ is
\begin{equation}
f(r^{\prime })=k+\frac{r^{\prime 2}}{2(n-2)(n-3)\alpha }\left( 1\pm \sqrt{1+%
\frac{4(n-2)(n-3)\alpha m}{r^{\prime n-4}}-\frac{4(n-2)(n-3)^{2}\alpha q^{2}%
}{(n-1)r^{\prime 2n-6}}}\right) .  \label{Fgn}
\end{equation}
Equation (\ref{Fgn}) for large values of $r^{\prime }$ becomes:
\begin{equation}
f_{\infty }(r^{\prime })=k+\frac{r^{\prime 2}}{4\alpha }(1\pm 1)\pm \frac{m}{%
r^{\prime (n-2)}}\mp \frac{(n-3)q^{2}}{(n-1)r^{\prime (2n-4)}},
\label{Fg1inf}
\end{equation}
which shows that the geometrical mass of the spacetime is the same as that
of Einstein gravity.

One should note that the function $f(r^{\prime })$ in Eq.
(\ref{Fg1}) is imaginary for $r^{\prime }<r_{0}$ and real for
$r^{\prime }>r_{0}$, where $r_{0}$ is the largest real solution of
\begin{equation}
3r_{0}^{6}+24\alpha (m+2\left| k\right| \alpha )r_{0}^{2}-8\alpha q^{2}=0.
\label{Eqr0}
\end{equation}
Of course, one may note that Eq. (\ref{Eqr0}) has real solution provided $%
\alpha >0$, or $32\alpha (m+2\left| k\right| \alpha )^{3}+3q^{4}<0$ for
negative $\alpha $. Thus, one cannot extend the spacetime to the region $%
r^{\prime }<r_{0}$. To get rid of this incorrect extension, we introduce the
new radial coordinate $r$ as
\begin{equation}
r^{2}=r^{\prime 2}-r_{0}^{2}\Rightarrow dr^{\prime 2}=\frac{r^{2}}{%
r^{2}+r_{0}^{2}}dr^{2}.  \label{trans}
\end{equation}
With this new coordinate, the metric (\ref{Met1}) and (\ref{Fg1}) become
\begin{eqnarray}
ds^{2} &=&-f(r)dt^{2}+\frac{r^{2}}{r^{2}+r_{0}^{2}}\frac{dr^{2}}{f(r)}%
+(r^{2}+r_{0}^{2})d\Omega ^{2},  \label{Met2} \\
f(r) &=&k+\frac{r^{2}+r_{0}^{2}}{4\alpha }\pm \sqrt{\left( \frac{%
r^{2}+r_{0}^{2}}{4\alpha }\right) ^{2}+\left( \left| k\right| +\frac{m}{%
2\alpha }\right) -\frac{q^{2}}{6\alpha (r^{2}+r_{0}^{2})}},  \label{Fg2}
\end{eqnarray}
and the vector potential is $A_{\mu }=-q/(2\sqrt{r^{2}+r_{+}^{2}})\delta
_{\mu }^{0}$. Of course, one may ask for the completeness of the spacetime
with $r\geq 0$. It is easy to see that the spacetime described by Eq. (\ref
{Met2}) is both null and timelike geodesically complete for $r\geq 0$ \cite
{Deh2}.

In order to study the general structure of these solutions, we first look
for the curvature singularities. It is easy to show that the Kretschmann
scalar $R_{\mu \nu \lambda \kappa }R^{\mu \nu \lambda \kappa }$ diverges at $%
r=0$ and therefore there is an essential singularity located at $r=0$. As
one can see from Eq. (\ref{Fg2}), the solution has two branches with ``$-$''
and ``$+$'' signs. We discuss them in the following subsections.

\subsection{Asymptotically de Sitter solutions}

We first investigate the ``$+$'' sign branch of $f(r)$ in Eq.
(\ref{Fg2}) with $k=1$. \ In this case $\alpha $ should be
nonzero, but can have negative or positive values. The negative GB
coefficient has been considered recently \cite{Deh3, Cor}. If
$\alpha <0$, then the metric of Eqs. (\ref {Met2}) and (\ref{Fg2})
has two inner and outer horizons located at
\begin{equation}
r_{\pm }=\left\{ \frac{1}{6}\left[ 3m\pm \sqrt{9m^{2}-12q^{2}}\right]
-r_{0}^{2}\right\} ^{1/2},
\end{equation}
provided $9m^{2}>12q^{2}$. Thus, one encounters with an asymptoticalle dS
black hole.

One can show that the Ricci scalar of the spacetime is $10/|\alpha
|$ as $r$ goes to infinity, and therefore the spacetime is
asymptotically de Sitter. Thus, in the Gauss-Bonnet gravity, one
can have asymptotically de Sitter black hole without any need to a
cosmological constant term in the field equations. It is
remarkable to note that there exist no asymptotically de Sitter
solutions for $k=0$, and $-1$.

\subsection{Asymptotically anti-de Sitter solutions}

Now, we consider the metric of Eqs. (\ref{Met2}) and (\ref{Fg2}) for the ``$%
+ $'' sign branch of $f(r)$ with positive values of $\alpha $ . The Ricci
scalar of the solution is $-10/\alpha $, and therefore the spacetime is
asymptotically anti-de Sitter. For the case of $k=0$ and $1$; $f(r)>0$ for
all values of $0\leq r<\infty $, and therefore this metric presents a naked
singularity. While for the case of $k=-1$, the function $f(r)$ in Eq. (\ref
{Fg2}) has a zero at $r=\{1/6[\sqrt{9m^{2}+12q^{2}}-3m]-r_{0}^{2}\}^{1/2}$,
and therefore we have an asymptotically AdS topological black hole. \

Note that there is no asymptotically anti-de Sitter nontopological black
hole in Gauss-Bonnet gravity without a cosmological constant. This feature
is different from the case of Einstein or Gauss-Bonnet gravity with
cosmological constant, which one has an asymptotically anti-de Sitter black
hole in the latter cases.

\subsection{Asymptotically flat solutions}

Now we discuss the branch of $f(r)$ in Eq. (\ref{Fg2}) with ``$-$'' sign.
One may note that for $k=0$, the function $f(r)$ goes to $0$ as $r$ goes to
infinity and therefore it is not acceptable. For $k=1$, one may note that $%
f(r)\rightarrow 1$ as $r$ goes to infinity and therefore the metric (\ref
{Met2}) and (\ref{Fg2}) is asymptotically flat. In this case for $\alpha >0$%
, the metric of Eqs. (\ref{Met2}) and (\ref{Fg2}) has two inner and outer
horizons located at $r_{-}$ and $r_{+}$, provided $9m^{2}>12q^{2}$. In the
case that $9m=12q^{2}$, we will have an extreme black hole, and for the case
of $9m^{2}<12q^{2}$, one encounters with a naked singularity. It is
remarkable to note that for large values of $r$, the function $f(r)$ can be
written as:
\begin{equation}
f(r)\simeq 1-\frac{m+2\alpha }{3r^{2}}+\frac{q^{2}}{3r^{4}},
\end{equation}
which shows that the spacetime behaves like a Reissner-Nordstrum black hole
with mass parameter $(m+2\alpha )/3$. Thus, one may conclude that the mass
of asymptotically flat black holes in Gauss-Bonnet gravity is more than the
mass of asymptotically flat black holes in Einstein gravity. For negative
values of $\alpha $, the metric presents a naked singularity, and in the
limit of $\alpha \rightarrow 0$, the spacetime is exactly asymptotically
flat Reissner-Nordstrum spacetime as one expected. For $k=$ $-1$ and $\alpha
<0$, this branch of $f(r)$ presents an asymptotically flat spacetime with
naked sinqularity and a cosmological horizon.

\section{Rotating Solutions}

\label{Rot}

Here we consider two classes of rotating solutions in Gauss-Bonnet gravity
without a cosmological constant for which the hypersurface of constant $r$
and $t$ are flat. As we have seen in the last section, one can only have
asymptotically AdS static solutions of these types.

\subsection{Charged rotating solutions}

First, we endow our spacetime solution (\ref{Met2}) and (\ref{flat}) with a
global rotation. The rotation group in $(n+1)$-dimensions is $SO(n)$ and
therefore the number of independent rotation parameters for a localized
object is equal to the number of Casimir operators, which is $[n/2]$, where $%
[z]$ is the integer part of $z$. Therefore for the case of a
five-dimensional spacetime, one can have at most two rotation parameters. It
is easy to show that the metric (\ref{Met2}) and (\ref{flat}) with two
rotation parameters $a_{1}$ and $a_{2}$ can be written as \cite{Deh1}
\begin{eqnarray}
ds^{2} &=&-f(r)\left( \Xi dt-{{\sum_{i=1}^{2}}}a_{i}d\phi _{i}\right) ^{2}+%
\frac{r^{2}}{r^{2}+r_{0}^{2}}\frac{dr^{2}}{f(r)}  \nonumber \\
&&\frac{r^{2}+r_{0}^{2}}{\mathbf{\alpha }}\left\{ \alpha ^{-1}{{%
\sum_{i=1}^{2}}}\left( a_{i}dt-\Xi \alpha d\phi _{i}\right) ^{2}-(a_{1}d\phi
_{2}-a_{2}d\phi _{1})^{2}+dx^{2}\right\} ,  \nonumber \\
\Xi ^{2} &=&1+\alpha ^{-1}{{\sum_{i=1}^{2}}}a_{i}^{2},\text{ \ \ \ \ \ }
\nonumber \\
A_{\mu } &=&\frac{q}{2r^{2}}((\Xi \delta _{\mu }^{0}-a_{i}\delta _{\mu
}^{i}),\hspace{1cm}\text{\textrm{(no sum on} }\mathrm{i)},
\end{eqnarray}
where the functions $f(r)$ is given by Eq. (\ref{Fg2}).

\subsection{Magnetic rotating solutions}

Here we want to obtain the $5$-dimensional solutions of Eqs. (\ref{Geq})-(%
\ref{Str}) which produce a longitudinal magnetic field normal to the $%
(r-\phi _{1})$-plane. We assume that the metric has the following form:
\begin{equation}
ds^{2}=-\frac{\rho ^{2}}{\alpha }dt^{2}+\frac{dr^{2}}{f(r)}+\alpha f(r)d\phi
_{1}^{2}+r^{2}d\phi _{2}^{2}+\frac{r^{2}}{\alpha }dx^{2}.  \label{Met3}
\end{equation}
Again, the coordinates $x$ have the dimension of length, while the angular
coordinate $\phi _{i}$ are dimensionless as usual and ranges in $0\leq \phi
_{i}<2\pi $. The motivation for this metric gauge $[g_{tt}\varpropto -r^{2}$
and $(g_{rr})^{-1}\varpropto g_{\phi _{1}\phi _{1}}]$ instead of the usual
Schwarzschild gauge $[(g_{rr})^{-1}\varpropto g_{tt}$ and $g_{\phi \phi
}\varpropto r^{2}]$ comes from the fact that we are looking for a magnetic
solution instead of an electric one. First, we consider only the static
solution. Since, we want to have a magnetic field, one may assume that $%
A_{\mu }=h(r)\delta _{\mu }^{\phi _{1}}$. Using the field equations (\ref
{Geq})-(\ref{Str}), one obtains
\begin{eqnarray}
f(r) &=&\frac{r^{\prime 2}}{4\alpha }\pm \sqrt{\frac{r^{4}}{16\alpha ^{2}}+%
\frac{m}{6\alpha }+\frac{q^{2}}{6\alpha r^{2}}},  \label{Fg3} \\
A_{\mu } &=&\frac{q}{2\sqrt{\alpha }r^{2}}\delta _{\mu }^{\phi _{1}}.
\label{Pot3}
\end{eqnarray}
The only nonvanishing component of electromagnetic field is $F_{r\phi _{1}}=%
\sqrt{\alpha }q/r^{3}$, which is a longitudinal magnetic field normal to the
$(r-\phi _{1})$-plane. In order to study the general structure of these
solutions, we first look for curvature singularities. It is easy to show
that the Kretschmann scalar $R_{\mu \nu \lambda \kappa }R^{\mu \nu \lambda
\kappa }$ diverges at $r=0$ and therefore there is an essential singularity
located at $r=0$. As one can see from Eq. (\ref{Fg3}), the solution has two
branches with ``$-$'' and ``$+$'' signs. Since the `` $-$'' signs branch
goes to zero as $r$ goes to infinity, therefore it cannot be accepted. The ``%
$+$'' signs branch is always positive, and therefore this spacetime has no
horizon. Thus, the metric (\ref{Met3})-(\ref{Pot3}) presents a naked
singularity.

Now we consider the most general magnetic rotating solution which can have
two rotation parameter in five dimensions. It is easy to show that the
following metric satisfies the field equations (\ref{Geq})-(\ref{Str}):

\begin{eqnarray}
ds^{2} &=&-\frac{r^{2}}{\alpha }\left( \Xi dt-{{\sum_{i=1}^{2}}}a_{i}d\phi
_{i}\right) ^{2}+\frac{f(r)}{\Xi ^{2}-1}{{\sum_{i=1}^{2}}}\left( (\Xi
^{2}-1)dt-\Xi \sum_{i=1}^{2}a_{i}d\phi _{i}\right) ^{2}  \nonumber \\
&&-\frac{r^{2}}{\alpha (\Xi ^{2}-1)}(a_{1}d\phi _{2}-a_{2}d\phi _{1})^{2}+%
\frac{dr^{2}}{f(r)}+\frac{r^{2}}{\alpha }dx^{2},  \nonumber \\
\Xi ^{2} &=&\alpha ^{-1}{{\sum_{i=1}^{2}}}a_{i}^{2},  \nonumber \\
A_{\mu } &=&\frac{q}{2r^{2}}((\Xi \delta _{\mu }^{0}-a_{i}\delta _{\mu
}^{i}),\hspace{1cm}\text{\textrm{(no sum on} }\mathrm{i)}.  \label{Met4}
\end{eqnarray}

\section{CLOSING REMARKS}

In this paper we investigated the classical theory of gravity without
cosmological constant. Indeed, we added the Gauss-Bonnet term to the
Einstein action and introduced a few solutions of the field equations in the
presence of an electromagnetic field. We found that one can have
asymptotically de Sitter, flat or anti-de sitter solutions in Gauss-Bonnet
gravity without any need to a cosmological constant term in gravitational
field equations. First, we introduced static solutions whose $3$-surfaces at
fixed $r$ and $t$ have constant positive ($k=1$), negative $(k=-1)$, or zero
curvature $(k=0)$. We encountered with two different branches for $f(r)$ in
Eq. (\ref{Fg2}). For ``$+$'' sign branch, we showed that when $k=1$, then
one could have asymptotically dS and AdS solutions. Indeed, for negative $%
\alpha $, the solution presented an asymptotically dS black hole with event
(E) and cosmological (C) horizons (H) provided $9m^{2}>12q^{2}$, while for
positive values of $\alpha $, the spacetime was asymptotically AdS with a
naked singularity (NS). For the case of zero $k$, one could have only
asymptotically AdS solutions, while for $k=-1$ and positive $\alpha $, one
had an asymptotically AdS topological black hole. For ``$-$'' sign branch,
the solutions were asymptotically flat. See for more details the following
table:

\bigskip

\begin{tabular}{|c|c|c|c|c|}
\hline
\begin{tabular}{l}
\textbf{Branch sign} \\
\ \ \ \textbf{of f(r)}
\end{tabular}
&  \ \ \textbf{k \ } & $\ \ \mathbf{\alpha }$ \ \  &
\begin{tabular}{l}
\textbf{Asymptotic} \\
\ \textbf{behavior}
\end{tabular}
& \textbf{Singularity} \\ \hline
$+$ & $1$ & $-$ & dS & BH with E\&CH ($9m^{2}>12q^{2}$) \\ \hline
$+$ & $1$ & $+$ & AdS & NS \\ \hline
$+$ & $0$ & $+$ & AdS & NS \\ \hline
$+$ & $-1$ & $+$ & AdS & BH with EH \\ \hline
$-$ & $1$ & $+$ & flat & BH with 2 H's ($9m^{2}>12q^{2}$) \\ \hline
$-$ & $1$ & $+$ & flat & BH with EH ($9m^{2}=12q^{2}$) \\ \hline
$-$ & $1$ & $+$ & flat & NS ($9m^{2}<12q^{2}$) \\ \hline
$-$ & $1$ & $-$ & flat & NS \\ \hline
$-$ & $-1$ & $-$ & flat & NS with CH \\ \hline
\end{tabular}

\bigskip

We found that the geometrical mass of these $5$-dimensional spacetimes is $%
m+2\alpha \left| k\right| $, which is different from the geometrical mass, $%
m $, of the solutions of Einstein gravity. It seems resonable to say that
the Gauss-Bonnet term with negative coefficient $\alpha $, acts as a
negative mass or antigravity effect. This feature occured only for the $5$%
-dimensional solutions, and was not repeated for the solutions of
Gauss-Bonnet gravity in higher dimensions.

We also added angular momentum to the static solutions with $k=0$, and
introduced the asymptotically AdS charged rotating solutions of Gauss-Bonnet
gravity with two rotation parameters. Finally, we found a class of solutions
which yields an asymptotically AdS spacetime with a longitudinal magnetic
field [the only nonzero component of the electromagnetic field is $F_{r\phi
} $] generated by a static magnetic brane. We found that these solutions
have curvature singularity at $r=0$ without any horizons. We also introduced
the magnetic rotating solutions with two rotation parameters. In these
spacetimes, when all the rotation parameters are zero (static case), the
electric field vanishes, and therefore the brane has no net electric charge.
For the spinning brane, when one or more rotation parameters are nonzero,
the brane has a net electric charge density which is proportional to the
magnitude of the rotation parameter given by $\sqrt{\Xi ^{2}-1}$.

As stated before, the Gauss-Bonnet gravity is the most general gravitational
field equation in five dimension. In higher dimension one should use more
terms for action in Lovelock theory. The consideration of asymptotically
(A)dS solutions in Lovelock gravity with more curvature terms remains to be
carried out in future.

\end{document}